\documentstyle[11pt,twoside,epsf,jltp]{article}

\title{Sliding Dynamics of the Wigner Crystal\\ on Liquid He}

\author{Yoshiyuki Shibayama, Masaya Yamazaki, Hanako Isshiki,\\ and Keiya Shirahama\address{Department of Physics, Keio University, Yokohama 223-8522, JAPAN}}

\runninghead{Y. Shibayama {\it et al.}}{Sliding Dynamics of the Wigner Crystal on Liquid He}

\begin{document}

\begin{abstract}

The Wigner crystal on liquid He accompanies
with periodic corrugation of the He surface;
dimples.
The dynamics of the crystal is coupled
with the motion and the deformation of the dimples.
Nonlinear phenomena found in AC Corbino conductivity
are attributed to the collective sliding of the electrons out of the dimples.
In order to inspect the dynamical transition to the sliding state,
we have developed a novel experimental method
using a so-called "$t^{2}$~pulse",
whose leading and trailing edges change in proportion to the square of time;
$V \propto t^{2}$.
Since the force exerting upon the crystal is proportional
to the time derivative of the input voltage,
$dV/dt$,
the $t^{2}$-pulsed method is expected to realize a continuous sweep
of the driving force,
resulting in the real-time observation of the sliding transition.
The observed response shows clearly the sliding,
revealing that the external force to the crystal
determines the sliding transition.

\end{abstract}

\maketitle

%Include this space if you do not use sections in your document.
%\vspace{0.3in}

\section{INTRODUCTION}
Surface state electrons on liquid He
form the Wigner crystal at low temperatures,
accompanying with periodic surface deformation,
referred to as {\it dimples}.
The Wigner crystal with the dimples has been discussed
in terms of the coupled plasmon-ripplon (CPR) model,~\cite{CPR}
and its first observation has been carried out
by the measurement of the CPR resonance.~\cite{CPRR}
The dynamics of the Wigner crystal is governed
by the motion and the deformation of the dimples,
because each of the electrons composing the Wigner crystal
is located in a potential caused by each of the dimples.

Recent studies of AC Corbino conductivity $\sigma_{xx}$ of the Wigner crystal
reveals an abrupt jump of $\sigma_{xx}$ at a certain driving voltage
and at a moderate magnetic field.~\cite{PRL,JLTP}
This conductivity jump is attributed
to the collective sliding of the electrons out of the potential.
In the $\sigma_{xx}$ measurement,
a sinusoidal voltage is applied to the inner Corbino electrode
coupled to the surface electrons capacitively,
and the resulting current is measured via the outer electrode
with a lock-in amplifier.
The signal averaging makes the detailed investigation of the sliding dynamics
difficult.

A transport study using a voltage pulse,
on the other hand,
makes real-time observation of the sliding achievable.
The measurement with a {\it trapezoidal} pulse~\cite{LT}
elucidates that the dynamics of the Wigner crystal is subjected
by the ramp rate of the input voltage,
$v=dV/dt$.
This is understood in terms of the sliding model.~\cite{PRL, JLTP}
In the present conditions,
$\sigma_{xx}$ is so large
that the current inside of the Wigner crystal is dominated
by the impedance of the capacitance
between the surface electrons and the Corbino electrode;
hence,
the current density $j$ at the electrode gap is proportional to $dV/dt$.
Since the external driving force to each of electrons
is given by $e{\sigma_{xx}}^{-1}j$,
the driving force is proportional to $dV/dt$.
Although this model does not take into account the current inhomogeneity
inside the surface electron,
it gives a basis for understanding
the nonlinear dynamics of the Wigner crystal.

In the trapezoidal pulse study,
$dV/dt$ change discontinuously at the edges of the pulse.
This means
that the Wigner crystal receives an impulsive force at the pulse edges.
It is therefore inappropriate to employ the trapezoidal pulse
to inspect sliding mechanism in detail.
The continuous sweep of the driving force
is obviously needed to investigate the dynamics of the transition
from the dimpled state to the sliding one and vice versa.

In the present work,
we have developed a novel experimental method
using a so-called "$t^{2}$~{\it pulse}",
in which the leading and trailing edges of the applied voltage pulse
are proportional to the {\it square} of time,
$V \propto t^2$.
Employing the $t^{2}$~pulse,
the ramp rate changes as $dV/dt\propto t$,
resulting in the continuous sweep of the driving force.
Consequently,
it is expected that the sliding transition is induced
in sweeping the driving force,
leading to real-time observation of the sliding dynamics.
In this paper,
we report the observation of the sliding transition
brought about by the $t^{2}$~pulse.
We have revealed that the sliding transition
is governed by only the ramp rate.

\section{EXPERIMENTAL}
As well as the previous works,~\cite{PRL,JLTP}
we have employed the capacitive coupling method
with a Corbino electrode.
A voltage pulse is applied to the inner electrode,
and the response current in the radial direction
is measured as an induced voltage at the outer electrode against the ground.
The typical waveform of the applied $t^{2}$~pulse
is shown in the inset of Fig.~1(a).
The input voltage is ramped up to a height $V_{0}$ with a rise time $t_{0}$,
where the voltage $V$ increases in proportion to the square of time;
$V\propto t^{2}$.
After the $t^{2}$~ramp up,
the voltage was kept at $V_{0}$ for 100~$\mu$s,
and then ramped down to zero with $V\propto t^{2}$ in a duration $t_{0}$.
This pulse is applied at every 200~$\mu$s.
The response is acquired by a digital storage oscilloscope
and is averaged 10000 times.

\begin{figure}
	\centerline{\epsfxsize= 4.4in\epsfbox{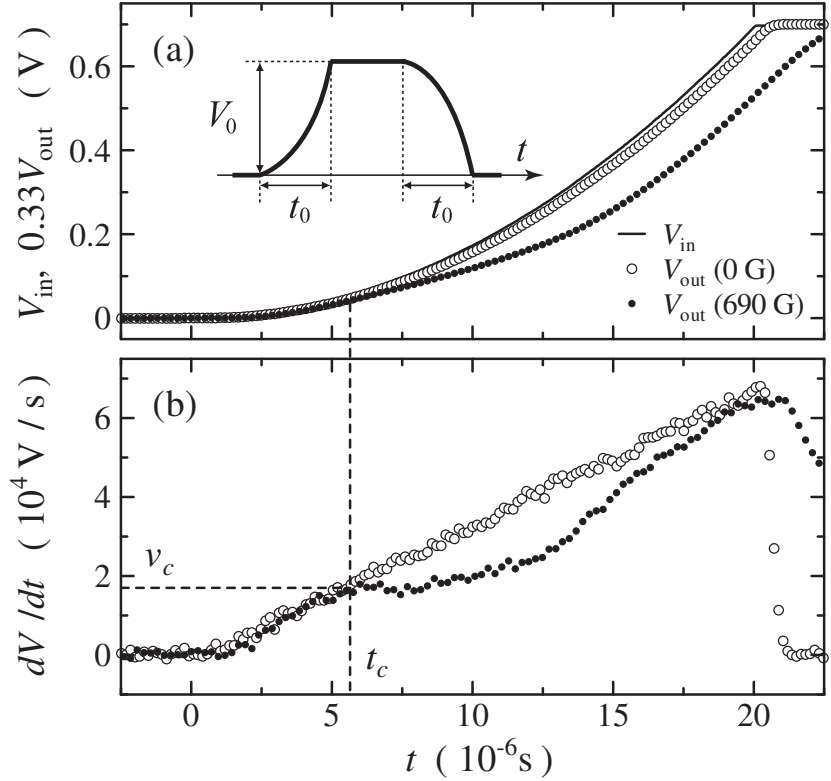}}
	\caption{
		(a)~The first half of the input $t^{2}$~pulse $V_{\rm in}$,
		and response $V_{\rm out}$ for $B=0$~G and 690~G,
		where the response is normalized to the height of $V_{\rm in}$.
		The inset shows the whole waveform of the $t^{2}$~pulse.
		(b)~The time derivative of the response.
		At 690~G,
		the sliding transition appears clearly at $t_{c}$,
		where the threshold ramp rate $v_{c}$ is defined
		as $(dV/dt)_{t=t_{c}}$.
	}
	\label{fig1}
\end{figure}

The electrons are generated at 1.4~K
by thermionic emission of a tungsten filament.
The present electron density is $1.52\times 10^{8}$~cm$^{-2}$.
The measurement is carried out at 60~mK,
much lower temperature than the melting point of the present Wigner crystal,
250~mK.\@
A static magnetic field $B$ up to 700~G
was applied perpendicular to the surface of the liquid He.

In order to compare the $t^{2}$-pulse results,
we have made the $\sigma_{xx}$ measurement using a 100~kHz sinusoidal wave
and have measured the responses for the trapezoidal voltage pulses.

\section{EXPERIMENTAL RESULTS AND DISCUSSION}
Figure 1(a) shows the first half of the input $t^{2}$~pulse $V_{\rm in}$
($V_{0}=700$~mV and $t_{0}=20$~$\mu$s)
and the response $V_{\rm out}$ for $B=0$~G and 690~G.\@
At 0~G,
the response follows the input pulse with a very short delay below 1~$\mu$s.
At 690~G,
on the other hand,
the response shows remarkably long delay
at $t>t_{c}\sim 5.7$~$\mu$s.
This delay becomes more prominent
by taking the time derivative of the response,
$dV/dt$,
as shown in Fig.~1(b).
At 0~G,
$dV/dt$ increases linearly,
indicating that the response trails the input waveform.
At 690 G,
in contrast,
the response shows an anomaly,
that is,
$dV/dt$ deviates from the linear tendency with a large delay at $t>t_{c}$.

\begin{figure}
	\centerline{\epsfxsize= 4.3in\epsfbox{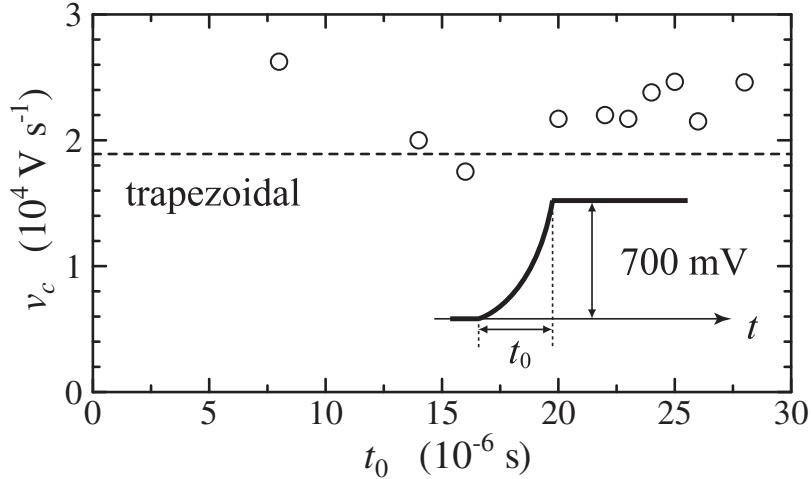}}
	\caption{
		The critical ramp rate $v_{c}$ as a function of the rise time $t_{0}$,
		where the pulse height $V_{0}$ is fixed at 700~mV
		as shown in the inset.
		The broken line denotes $v_{c}$
		derived from a trapezoidal pulse measurement.
	}
	\label{fig2}
\end{figure}

Our previous studies on the nonlinear transport of the Wigner crystal
employing both continuous waves (CW)~\cite{PRL,JLTP}
and trapezoidal pulses~\cite{LT} illustrate
that the sliding transition is brought about
when the ramp rate of the applied voltage exceeds a certain critical rate.
In the trapezoidal-pulsed study,
moreover,
the sliding is observed as a sudden increase of the response delay.
Therefore,
the present behavior of the $t^{2}$-response waveform at $t>t_{c}$
means the sliding transition of the Wigner crystal
induced by the $t^{2}$~pulse.
Based upon the data shown in Fig.~1,
we determine the critical ramp rate
$v_{c}\equiv (dV/dt)_{t=t_{c}}=1.8\times 10^{4}$~V~s$^{-1}$.

\begin{figure}
	\centerline{\epsfxsize= 4.4in\epsfbox{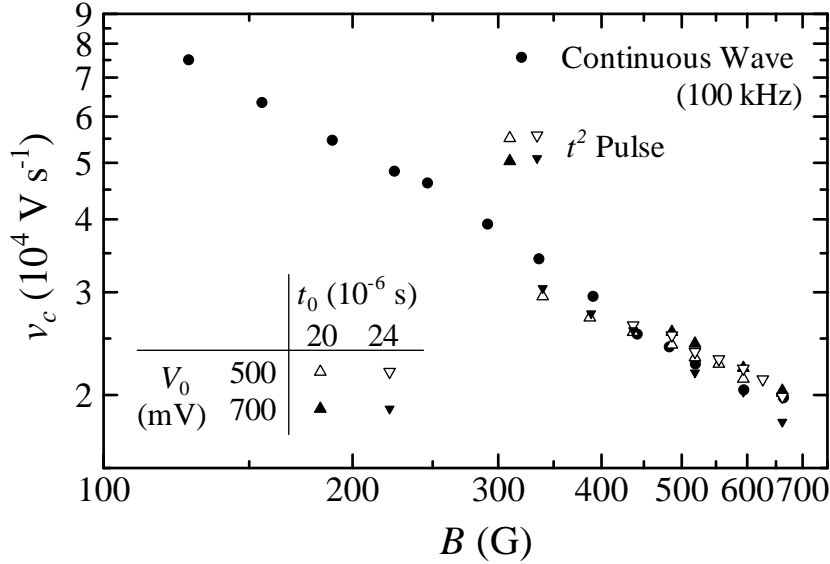}}
	\caption{
		$v_{c}$ as a function of the applied magnetic field $B$.
		Closed circles denote $v_{c}$ obtained from the CW measurement.
		Others are obtained from $t^{2}$~responses,
		where $t_{0}$ and $V_{0}$ are shown in the inset.
	}
	\label{fig3}
\end{figure}

We have made the above measurements for various $t_{0}$'s.
In Fig.~2,
$v_{c}$ is shown as a function of $t_{0}$
for a fixed $V_{0}=700$~mV.\@
The critical ramp rate obtained from the trapezoidal pulse measurement
is also shown as a broken line.
In the $t^{2}$-pulsed method,
$v_{c}$ is irrespective of $t_{0}$,
although it ranges from 2.0 to $2.5\times 10^{4}$~V~s$^{-1}$.
This indicates that the sliding transition
is not subjected by the rise time,
but by the ramp rate of the input pulse.
Moreover,
$v_{c}$ is fairly close to the one obtained
in the trapezoidal pulse measurement,
$1.9\times 10^{4}$ V s$^{-1}$.
This agreement convinces
that the sliding state caused by the $t^{2}$~pulse
is the same state as the one induced by the trapezoidal pulse.

We have found
that $v_{c}$ obtained from the $t^{2}$~pulse is always slightly larger
than that in the trapezoidal method.
This may be substantial,
because the sliding transition can be dominated
not only by the voltage ramp rate
but also by the history in the applied voltage.
In short,
the impulsive force produced at the edges of the trapezoidal pulse
may activate the sliding.
This proposition must be clarified
by the detailed analysis of the response waveform
below and above $v_{c}$,
and it will be discussed elsewhere.

Figure 3 shows the magnetic field dependence of $v_{c}$
for various pulse conditions,
together with the $v_{c}$ data obtained
from the $\sigma_{xx}$ jumps in the 100~kHz continuous wave measurement.
All the $v_{c}$ data collapse onto a single line in the log-log plot,
showing the powerlaw behavior:
$v_{c}\propto B^{-0.73}$.
This behavior is consistent with the result in our previous work,
where the powerlaw is explained by the transition
from the CPR state to a non-CPR state.~\cite{PRL,JLTP}
As regards the $t^{2}$-pulsed studies,
$v_{c}$ is in good agreement
with the one obtained from the $\sigma_{xx}$ jump.
Again this fact shows
that the present transition brought about by the $t^{2}$ pulse
is the same phenomenon caused by the continuous wave.
In the $t^{2}$-pulsed studies,
$v_{c}$ is not obtained below 300~G,
because the delay of the response has not been observed.
The reason has not yet been clarified.
We suppose that the disappearance
is not intrinsic property of the pulsed method,
because the sliding transition has been verified
in the trapezoidal pulse response down to 200~G.~\cite{LT}\@
Employing the $t^{2}$ pulses with different $V_{0}$ and/or $t_{0}$
may make the observation of the sliding transition at the lower magnetic field
possible.

\section{CONCLUSION}
In order to inspect the sliding dynamics of the Wigner crystal in real-time,
the pulsed method employing a $t^{2}$~pulse has been developed,
where the exerting force upon the Wigner crystal is swept continuously.
The sliding transition of the crystal is induced by the $t^{2}$~pulse
as well as by a trapezoidal one.
The sliding is determined by the ramp rate of the input pulse.
As regards $B$ dependence of $v_{c}$,
the powerlaw behavior is observed,
which is in good agreement with our previous results and the sliding model.

As compared to the previous works,
the present study has the advantage of the continuous sweep
of the driving force to the Wigner crystal.
The continuous sweep gives prominence
to the transition from the dimpled to the sliding state.
The quantitative analysis of the response waveform below and above $v_{c}$
will clarify the sliding dynamics of the Wigner crystal.

\section*{ACKNOWLEDGMENTS}
This work is supported by Grant-in-Aid for Scientific Research (B)
from the Ministry of Education, Culture, Sports, Science and Technology
of Japan,
and by Yamada Science Foundation.

\end{document}